\documentclass[useAMS,usenatbib]{mn2e}
\usepackage{graphicx,amsmath,amssymb,subfigure}

\def\simgt{\mathrel{\lower0.6ex\hbox{$\buildrel {\textstyle >} \over {\scriptstyle \sim}$}}}
\def\simlt{\mathrel{\lower0.6ex\hbox{$\buildrel {\textstyle <} \over {\scriptstyle \sim}$}}}
\newcommand{\gtsim}{\mbox{{\raisebox{-0.4ex}{$\stackrel{>}{{\scriptstyle\sim}}$}}}}

\newcommand{\apj}{ApJ}
\newcommand{\apjl}{ApJL}
\newcommand{\mnras}{MNRAS}
\newcommand{\aj}{AJ}
\newcommand{\apjs}{ApJS}
\newcommand{\nat}{Nature}
\newcommand{\araa}{ARA\&A}
\newcommand{\aap}{A\&A}
\newcommand{\aaps}{A\&AS}
\newcommand{\pasp}{PASP}

\begin{document}

\title[A young, dusty, compact radio source within a large Ly$\alpha$
  halo]{A young, dusty, 
compact radio source within a Ly$\alpha$ halo}

\author[F.E.Barrio et al.]
{F.~Eugenio Barrio$^{1}$\thanks{Email: febm@astro.ox.ac.uk},
Matt J.~Jarvis$^{1,2}$\thanks{Email: m.j.jarvis@herts.ac.uk},
Steve Rawlings$^{1}$, \and Amanda Bauer$^{3,4}$, Steve
Croft$^{5,6,7}$, Gary J.~Hill$^{3}$, Arturo Manchado$^{8}$,
\and Ross J.~McLure$^{9}$, Daniel J.B.~Smith$^{1,10}$, Thomas A.~Targett$^{9}$ \\ \footnotesize
$^{1}$Astrophysics, Department of Physics, Keble Road, Oxford, OX1 3RH, UK \\
$^{2}$Centre for Astrophysics Research, Science \& Technology Research
Institute, 
University of Hertfordshire, Hatfield, AL10 9AB, UK\\
$^{3}$University of Texas at Austin, Austin, TX 78712, USA\\
$^{4}$Gemini Observatory, Southern Operations Center c/o AURA, Casilla 603, La Serena, Chile\\
$^{5}$Inst. of Geophysics and Planetary Physics, Lawrence Livermore
National Laboratory L-413, 700 East Ave, Livermore, CA 94550, USA\\
$^{6}$UC Davis, 1 Shields Ave, Davis, CA 95616, USA\\
$^{7}$UC Merced, P.O. Box 2039, Merced, CA 95344, USA\\
$^{8}$Instituto de Astrof\'isica de Canarias, c/ V\'ia L\'actea s/n,
38205, La Laguna, Tenerife, Espa\~na, affiliated to CSIC, Espa\~na\\
$^{9}$Institute for Astronomy, University of Edinburgh, Royal Observatory, Blackford Hill, Edinburgh, EH9 3HJ, UK\\
$^{10}$Astrophysics Research Institute, Liverpool John Moores University, Egerton Wharf, Birkenhead, CH41 1LD, UK
}

\maketitle

\begin{abstract}
%We are searching for radio-loud quasars within the epoch of
%reionization by selecting as candidates, flat-spectrum radio
%sources that are bright and unresolved in the near-infrared,
%but very faint optically.
We report here on the discovery of
a red quasar, J004929.4+351025.7 at a redshift of $z = 2.48$, situated 
within a large 
Ly$\alpha$ emission-line halo. The radio spectral energy 
distribution implies that the radio jets were triggered
$<10^4$~years prior to the time at which the object is
observed, suggesting that the jet triggering of the active
galactic nucleus is recent. The loosely biconical structure of
the emission-line halo suggests that it is ionised by photons
emitted by the central quasar nucleus and that the central nucleus is
obscured  by a dusty torus with $A_{\rm v} \sim 3.0$. The large spatial
extent of the Ly$\alpha$ halo relative to the radio
emission means this could only have occurred if the radio jets emerged from an
already established highly-accreting black hole. This suggests
that the radio-jet triggering is delayed with respect to the
onset of accretion activity on to the central supermassive
black hole.
\end{abstract}


\section{Introduction}

Most quasars exhibit a blue optical and ultraviolet
continuum, broad emission lines, and narrow forbidden emission lines of
highly ionised elements \citep[e.g.][]{2003AJ....126.2579S}.
However,
there is also a population of reddened quasars
predicted by the unification schemes \citep{1995PASP..107..803U},
and which have been known for a number of years
\citep[e.g.][]{1995Natur.375..469W}. The origin of red quasars and
their fraction relative to 
normal quasars is still
subject of debate: we still do not know the fraction that are
intrinsically red, e.g.\ in radio-loud quasars
the synchrotron tail can extend into the near-infrared and
optical wavebands \citep[][]{2001MNRAS.323..718W}, or the fraction that are
reddened by dust \citep[e.g.][]{2003AJ....126.1131R}. For the dusty
quasars, it is not clear if the dust is typically 
situated in a dusty torus around the central quasar nucleus
\citep[e.g.][]{1993ARA&A..31..473A}, or whether it is distributed more widely
throughout the host galaxy \citep[e.g.][]{2006MNRAS.370.1479M}.

Determining the
fraction of very red quasars (red, reddened or both) in the
total population is essential for a precise determination of
the quasar Luminosity Function (LF) and also for understanding the
evolution of quasars and the nature of the quasar-galaxy
connection. If a large fraction is eliminated from optical surveys
by simple dust reddening
then the relative comoving space density of such objects may be much
higher and would have a much more profound effect on the evolution of
massive galaxies. 

Historically optical surveys, relying on the
ultraviolet-excess and power-law shape of the typical (i.e.
blue) quasar spectrum, have failed to discover this red
population, principally due to the fact that the blue excess is not present
in reddened quasars \citep{2004ApJS..155..257R}. Unsurprisingly,
recent radio and infrared
surveys have shown a better promise in finding red quasars over all redshifts
\citep{2007ApJ...667..673G}. Although previous large
area attempts have suffered from incompleteness, this is still
the most promising approach.

Many powerful, high-redshift radio-loud active galactic
nuclei (AGN), are now known to be surrounded by large-scale,
highly-luminous, Ly$\alpha$ emission-line haloes 
\citep[][]{1988ApJ...327L..47C, 2002MNRAS.336..436V, 
2003ApJ...592..755R}. Similar Ly$\alpha$
emission haloes have also been found
%even when there is
around a radio-quiet quasar \citep{2005A&A...436..825W}, and
submillimetre galaxies \citep{2000ApJ...532..170S,
2004ApJ...606...85C, 2004MNRAS.351...63B, 2005MNRAS.363.1398G}.
This suggests that giant Ly$\alpha$ halos may be
%may be commonly
associated with the onset of quasar and/or starburst activity
\citep[see also][]{2001ApJ...556...87H, 2003ApJ...591L...9O,
2005Natur.436..227W}. However, the recent discovery of a number
of large diffuse Ly$\alpha$ haloes around seemingly quiescent
galaxies
\citep{2006A&A...452L..23N,2007astro.ph..3522S,2007arXiv0705.1494S}
suggests a model in which the origin of some Ly$\alpha$ haloes
is cold gas accreting on to the
dark-matter halo of a massive galaxy
\citep{2001ApJ...562..605F,2006ApJ...649...14D}. There may therefore be
several mechanisms responsible for giant
Ly$\alpha$ haloes. However, most of the large $>50$~kpc haloes discovered so far
seem to surround massive galaxies, thus the large haloes
may be a  product or a  requisite ingredient in the formation of
massive galaxies at 
high redshift, regardless of the mechanism behind their ionisation. 

In this paper we report the discovery of a reddened quasar that
conforms to the idea that 
red, young, radio quasars at $z \sim 2.5$ might pick out dusty,
possibly star-forming, massive galaxies within giant Ly$\alpha$ haloes.
%In section~\ref{sec:survey} we present the survey strategy.
In section~\ref{sec:data} we present the photometric and
spectroscopic data and in section~\ref{sec:discuss} we discuss
the properties of this object. In section~\ref{sec:conc} we
make some concluding remarks concerning the implications of
this object for jet-triggering in AGN. We assume throughout
that $H_{0} =70\,{\rm km\,s^{-1} Mpc^{-1},}\; \Omega_{\rm M} =
0.3$ and $\Omega_{\Lambda} = 0.7$. All quoted magnitudes are in
the Vega system.

\section{Data \& observations}\label{sec:data}

\begin{table}% TABLE 1 radio
\begin{center}
\caption{Summary of radio properties of J0049+3510. The
quoted limits are 5$\sigma$.}\label{T_radio}
\begin{tabular}{lrr@{ $\pm$ }l}
\hline
Survey & Frequency / &\multicolumn{2}{c}{Flux Density/}\\
&\multicolumn{1}{c}{GHz}&\multicolumn{2}{c}{\rm mJy}\\
\hline
VLA  &\hspace{1.5em}8.4&33.4&1.4\\
GB6  &4.85&67&9\\
NVSS &1.4&122.0&3.7\\
WENSS&0.325&101&18\\
6C   &0.151&\multicolumn{2}{c}{$<$ 400}\\
VLSS &0.074&\multicolumn{2}{c}{$<$ 500}\\
\hline
\end{tabular}
\end{center}
\end{table}

\subsection{Quasar discovery}
J004929.4+351025.7 (hereafter J0049+3510)
was
initially found by cross-matching the 325~MHz Westerbork
Northern Sky Survey \citep[WENSS;][]{1997A&AS..124..259R} and
the 1.4~GHz Northern VLA Sky Survey
\citep[NVSS;][]{1998AJ....115.1693C} radio source catalogues
as part of our survey for $z>6$ quasars \citep[][]{2004ASPC..311..361J}.
It has a 325~MHz flux density of $S_{325{\rm MHz}} = 101 \pm
18\,{\rm mJy}$ and a 1.4GHz flux-density of $122.0 \pm 3.7$~mJy,
giving an inverted radio spectrum between 325{\rm MHz} and
1.4{\rm GHz}, $\alpha^{1.4{\rm GHz}}_{325{\rm MHz}} = -0.134$.
As inspection of the Digitized Sky Survey showed no optical
counterpart at the radio position, it was marked for
follow up in the optical and near-infrared wavebands as a
candidate $z>6$ quasar.

\subsection{Radio data}\label{radio}
A search in the literature showed counterparts for
J0049+3510 in the GB6 4.85 {\rm GHz} survey
\citep{1996ApJS..103..427G} but no counterparts in the  6C
151 {\rm MHz} survey \citep{1993MNRAS.262.1057H}, nor the VLSS
74 {\rm MHz} survey \citep{2007arXiv0706.1191C}. The source
position is not covered by the FIRST survey
\citep{1995ApJ...450..559B}.

\begin{figure}%FIG 1 RADIO SED
\begin{center}
\includegraphics[height=0.38\textwidth, width=0.46\textwidth]{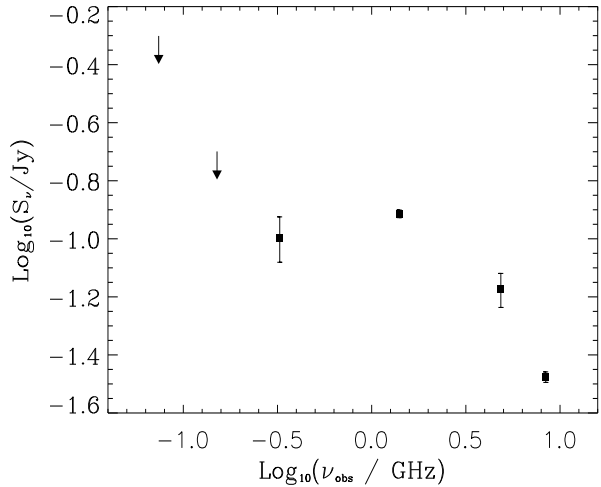}
\caption{Observed-frame radio spectral energy distribution for
  J0049+3510. The peak of the SED is shown to be around 1.2 GHz (4.2~GHz in
  the rest frame for $z \sim 2.5$), corresponding to a source age of $\sim
  10^{3}$~years \citep[see][]{1998PASP..110..493O}.}\label{F_SED}
\end{center}
\end{figure}

Archive observations from the VLA, in A-array, X-band ($8.0-8.8 {\rm
  GHz}$), were recovered and a map made from the data
available. This revealed a compact radio source (unresolved at
0.3 {\rm arcsec} resolution) with a flux density $S_{8{\rm GHz}} =
33.4\pm1.4$~mJy (see Table \ref{T_radio}). The radio-spectrum
of J0049+3510 is shown in Figure~\ref{F_SED}, and it is
apparent that the SED is peaked around 1.2~GHz. Given the compact
  nature and the peak of the SED at 1.2~GHz we classify this quasar as
  a Giga-Hertz-Peaked-Spectrum \citep[GPS;][]{1998PASP..110..493O}
  source. 

\begin{table*}%TABLE 2 observation log and photometry
\caption{Summary of optical and near-infrared imaging
observations and photometry of J0049+3510. Quoted limits
are 3$\sigma$. J0049+3510 is a point source in the $J$-,
$H$-, and $K$-band images.}\label{T_ima-log}
\begin{tabular}{c l c c c c c}
\hline
Date&\multicolumn{1}{c}{Telescope \&}   &Filter &Exposure   &Seeing     &Aperture   &Magnitude\\
    &\multicolumn{1}{c}{Instrument}     &       &(s)        &(arcsec)   &(arcsec)   &Vega     \\%(2 {\rm arcsec} aperture)\\
\hline
2004-12-13&INT--WFC         &$g$&1800     &1.1    &10   &19.56 $\pm$ 0.15       \\
2003-08-28&HJST-IGI       &$R$&150     &1.2    &2    &$>$ 23.9               \\
2003-09-22&WHT--PFIP        &$I$&1200     &1.2    &2    &$>$ 24.5               \\
2004-11-04&WHT--LIRIS       &$J$&3000     &0.6    &2    &21.06 $\pm$ 0.20       \\
2003-09-26&UKIRT--UFTI      &$H$&540      &0.7    &2    &18.69 $\pm$ 0.15       \\
2003-09-15&UKIRT--UFTI      &$K$&540      &0.5    &2    &17.62 $\pm$ 0.07       \\
\hline
\end{tabular}
\end{table*}

\subsection{Optical and near-infrared imaging}
The quasar was subsequently observed at Harlan J. Smith
Telescope (HJST) at McDonald Observatory for 150$\,$s in the
$R$-band.  The data were bias-subtracted, flat-fielded, and
combined with standard {\sc iraf} packages. For flux
calibration we used standard stars observed on the same night.
Aperture photometry was performed using the {\sc iraf apphot}
package with a 2~arcsec diameter aperture.  No identification
was found at the position of the radio source, implying that
our source was fainter than $R\simeq 23.9$~mag at the $3
\sigma$ level.

\begin{figure*}% FIGURE 2 OVERLAYS
\includegraphics[height=0.48\textwidth]{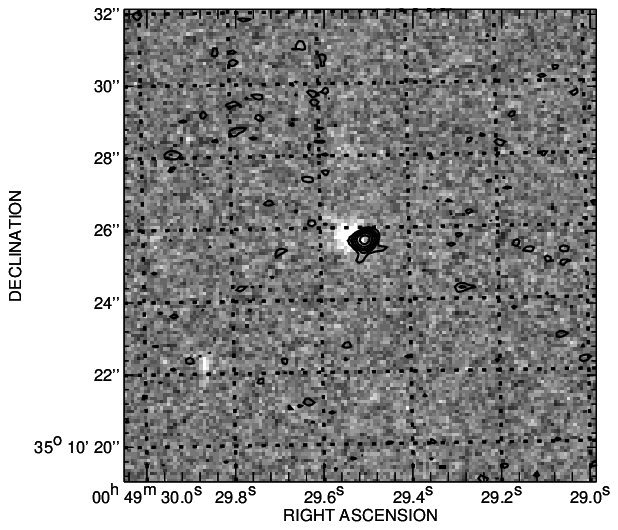}
\includegraphics[height=0.48\textwidth]{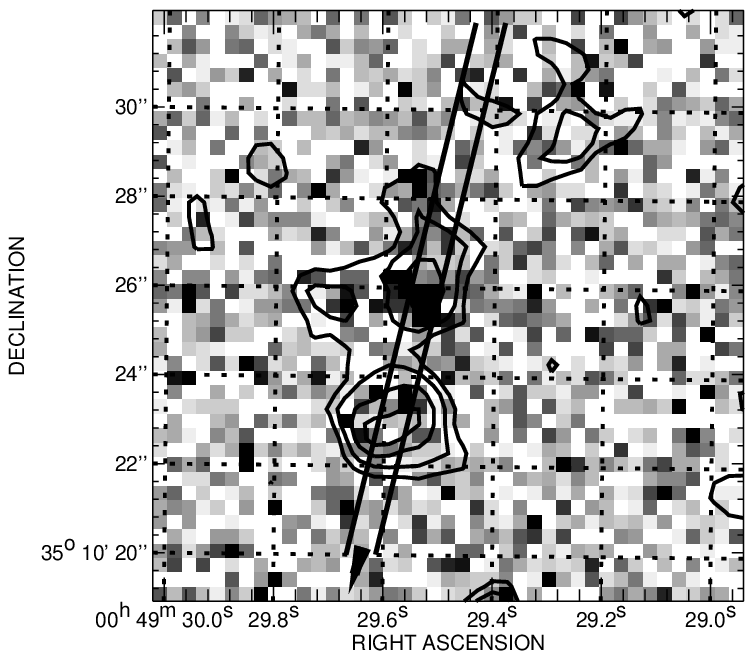}%, width=0.4\textwidth
\caption{({\it left}) $K$-band image (greyscale) overlayed with
  the radio $8 {\rm GHz}$ (contours) showing the compact radio
  source matching the unresolved near-infrared object within
  the astrometric uncertainty of the $K$-band image. The radio
  contours are $-2 \sigma$ and $2^{\frac{n}{2}} \times \sigma$,
  where n= 2, 3, 4, \ldots, 14  with $\sigma = 282\,\mu{\rm
  Jy}$. ({\it right}) The $K$-band 
  image (grey-scale) is overlayed with the optical $g$-band
  (contours) showing diffuse blue emission near to our source
and a   %($\sim xx \sigma$)
  detection of the quasar itself, which we called the south and
  central blobs respectively. There is also tentative evidence
  for further Ly$\alpha$ emission to the north-west
  of the central source. However, this is at a
  much lower level than the central and southern components.
  There are no co-spatial identifications for either the
  southern and northern components down to $K = 19$
  ($3\sigma$). We have also added the position of the long slit
  used in the WHT--ISIS spectroscopy.} \label{F_overlays}
\end{figure*}

Following our search method, we targeted it with near-infrared
$K$-band imaging on the UK Infrared Telescope (UKIRT)
with good seeing conditions ($< 0.6$ arcsec). The UKIRT
imaging was carried out with the UKIRT Fast-Track Imager (UFTI) using the
standard nine-point jitter pattern and reduced using {\sc oracdr}
pipeline, which subtracts the dark current and flat-fields the
data with a median sky-flat constructed from the individual
science frames. These observations showed an unresolved source
within 0.5~arcsec of the radio position with $K = 17.62 \pm
0.07$ (Fig.~\ref{F_overlays}).

J0049+3510 was then targeted in $J$- and $H$-bands, again,
with the UKIRT-UFTI using the same strategy as for the $K$-band
observations, which showed that it was also identified in the
$H$-band with $H=18.69 \pm 0.15$, however the $J$-band
observations were carried out in poor conditions and an
accurate $J$-band magnitude could not be obtained.

Subsequent deep optical $I$-band imaging on the
William Herschel Telescope (WHT) produced blank fields with
a limit of $I > 24.5$ ($3\sigma$). Deep $J$-band
observations were also carried out with the Long-slit
Intermediate Resolution Infrared Spectrograph (LIRIS) on the
WHT, these observations showed a point source with $J = 21.06
\pm 0.20$ in 0.6 $\rm arcsec$ seeing. A summary of the optical and
near-infrared 
photometry data is presented in Table \ref{T_ima-log}.

At this stage J0049+3510 was considered a good candidate
for a $z \geq 6$ quasar as the optical and near-infrared SED
suggested the presence of a steep spectral break between the
near-infrared and optical bands. This would correspond to the
Gunn-Peterson trough, with the Ly$\alpha$ line redshifted
between the $I$- and $J$-bands. We therefore sought
spectroscopy in order to determine the redshift and nature of
this source.

\begin{figure}% Fig 3 infrared spectroscopy 1D
\begin{center}
\includegraphics[angle=90, width=0.5\textwidth,]{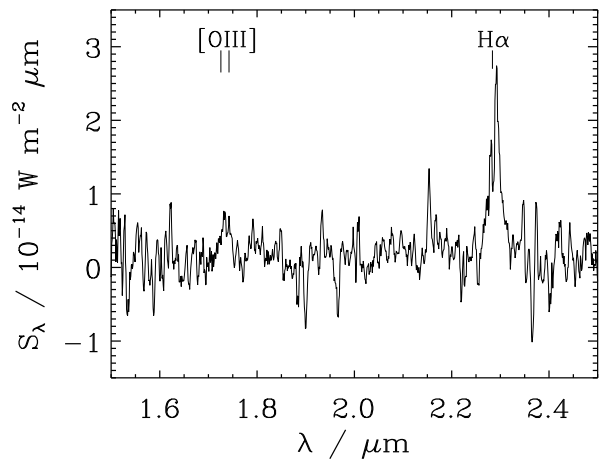}
\caption{Near-infrared $HK$ spectrum of quasar J0049+3510.
The spectrum has been smoothed using a $0.003\,\mu{\rm m}$ FWHM gaussian,
with the broad H$\alpha$ line at $2.292\,\mu{\rm m}$, and a very weak
[O {\sc iii}] doublet confirming the redshift of this source as
$z = 2.48$.} \label{F_ir-spec}
\end{center}
\end{figure}

\subsection{Optical and near-infrared spectroscopy}
Near-Infrared spectroscopy observations were performed with the
UKIRT Imaging Spectrometer (UIST) using the standard `ABBA'
nodding sequence (with a nod of 12~arcsec) on the night of
2004-09-16. The total exposure time was 160~minutes and the
data were reduced with {\sc oracdr}. The spectrum was extracted
over 1.2 arcsec (the pixel scale is $0.12\,{\rm arcsec \,
pixel}^{-1}$) with the {\sc iraf apextract} packages. We
identified a broad emission line at $2.292\,\mu{\rm m}$, first
identified, in view of the optical and near-infrared SED, as
the MgII$\lambda2799$ line redshifted to $z = 7.2$. No other
obvious lines were identified in the $HK$ Spectrum, but continuum
was present throughout the wavelength range. Figure
\ref{F_ir-spec} shows the extracted one-dimensional
near-infrared spectrum of the quasar. The spectrum also shows
tentative evidence for a second emission line at 1.74 $\mu$m
which we later confirmed to be the [O{\sc iii}] doublet at a
redshift of $z \sim 2.48$, thus suggesting that the bright
emission line may be H$\alpha$. If the redshift were
indeed $z = 2.48$ or $z=7.2$, then we should be able to detect the
Ly$\alpha$ line with optical spectroscopy.

Optical spectroscopy was carried out with the Gemini Multi--Object
 Spectrograph (GMOS) on Gemini North. We used a 0.75 {\rm arcsec} long
 slit with PA of 78$\,^{\circ}$ with a nodding of 10 {\rm arcsec}. We
 selected the G150 grating centred at 750 nm for a total exposure time
 of 1200~s. The data was reduced with {\sc iraf} and a spectrum was
 wavelength-, flux-calibrated and extracted over 1.5 {\rm arcsec} (the
 spatial scale was $0.145 \,{\rm arcsec \, pixel}^{-1}$ ) with our own
 {\sc idl} routines. The spectrum showed no lines but a very faint
 continuum roughly between 4800 and 8000 {\rm \AA}. The seeing,
 measured as the Full Width Half Maximum (FWHM) of a gaussian fit to
 the spatial profile of an unresolved object on the slit, was 1.3 {\rm
 arcsec}. 
%Spectroscopy at shorter wavelengths was needed to detect  the Ly$\alpha$ line. 
 %In such non-optimal conditions    %Using one other object on the
 %slit we were able to measure a relative flux and from a previous
 %image estimate a photometry yielding 21.96 mag ($1.66 \times 10^{-21}
 %{\rm W \, m^{-1} \, Hz^{-1}}$). 

\subsection{The Ly$\alpha$ Halo}
The quasar was observed using the Sloan Gunn $g$
($\approx 420-550 ~ \rm nm$) filter used with the Isaac Newton
Telescope Wide-field camera (INT--WFC) for 1800~s during the
night of 2003-12-13. The image showed extended diffuse
structure to the south of the quasar position in addition to
some emission co-spatial with the unresolved $K$-band
identification which we take to be dominated by line emission; we
refer to these as the south and central 
blobs respectively. Figure \ref{F_overlays} shows the $K$-band
image overlayed with the radio contours and also with the
contours representing the $g$-band image. There is also tentative evidence
for further Ly$\alpha$ emission towards the north of the source, which
is much fainter than the central and southern components. However, we
believe it to be part of the halo, although deeper observations are
needed to confirm this. 

Subsequent WHT--ISIS spectroscopy on 2004-12-09 targeted the
brightest two (central and south) blobs
using a 2~arcsec slit with position angle of $170^{\circ}$ east of
north, with a total exposure time of 40~min. We used the
R300B grating, giving a spectral resolution of $\sim 9$\AA. The ISIS spectra
were reduced with  {\sc iraf}, wavelength calibrated with a
CuAr+CuNe arc and flux calibrated using observations of the
SP1942+261 standard star. The one-dimensional spectrum was
extracted with the {\sc iraf apextract} task for several
different apertures.

\begin{figure*}% Fig 4 optical spectroscopy 2D + 1D over  2 regions
\begin{center}
\includegraphics[width=0.5\textwidth ]{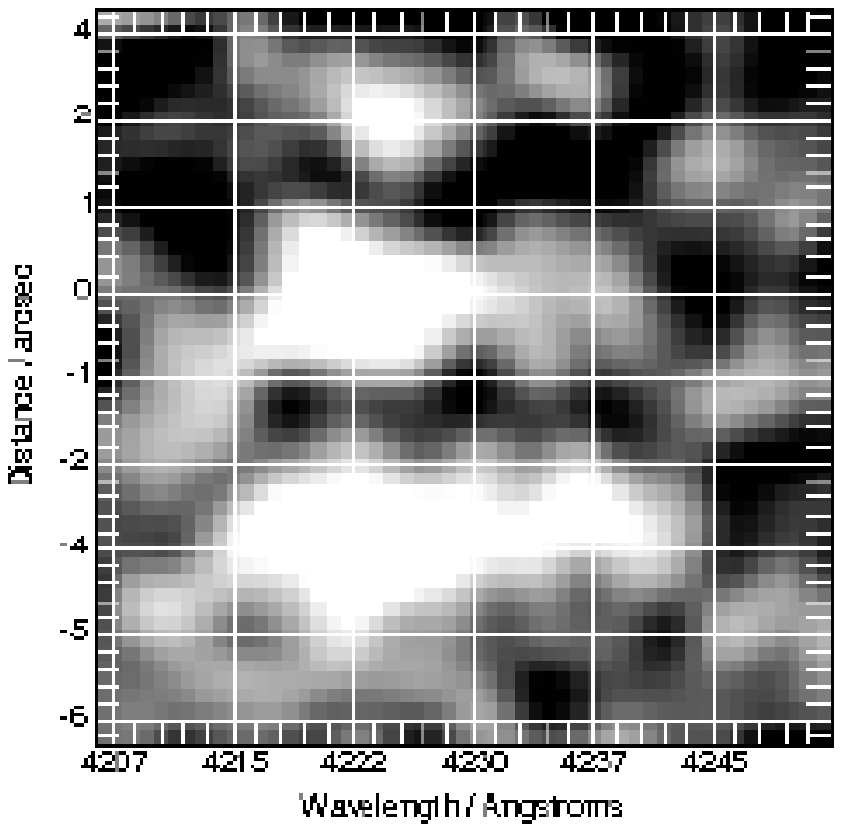} %height=0.55\textwidth %{2dspec.ps},height=0.45\textwidth,width=0.45\textwidth
\includegraphics[width=0.48\textwidth, height=0.47\textwidth]{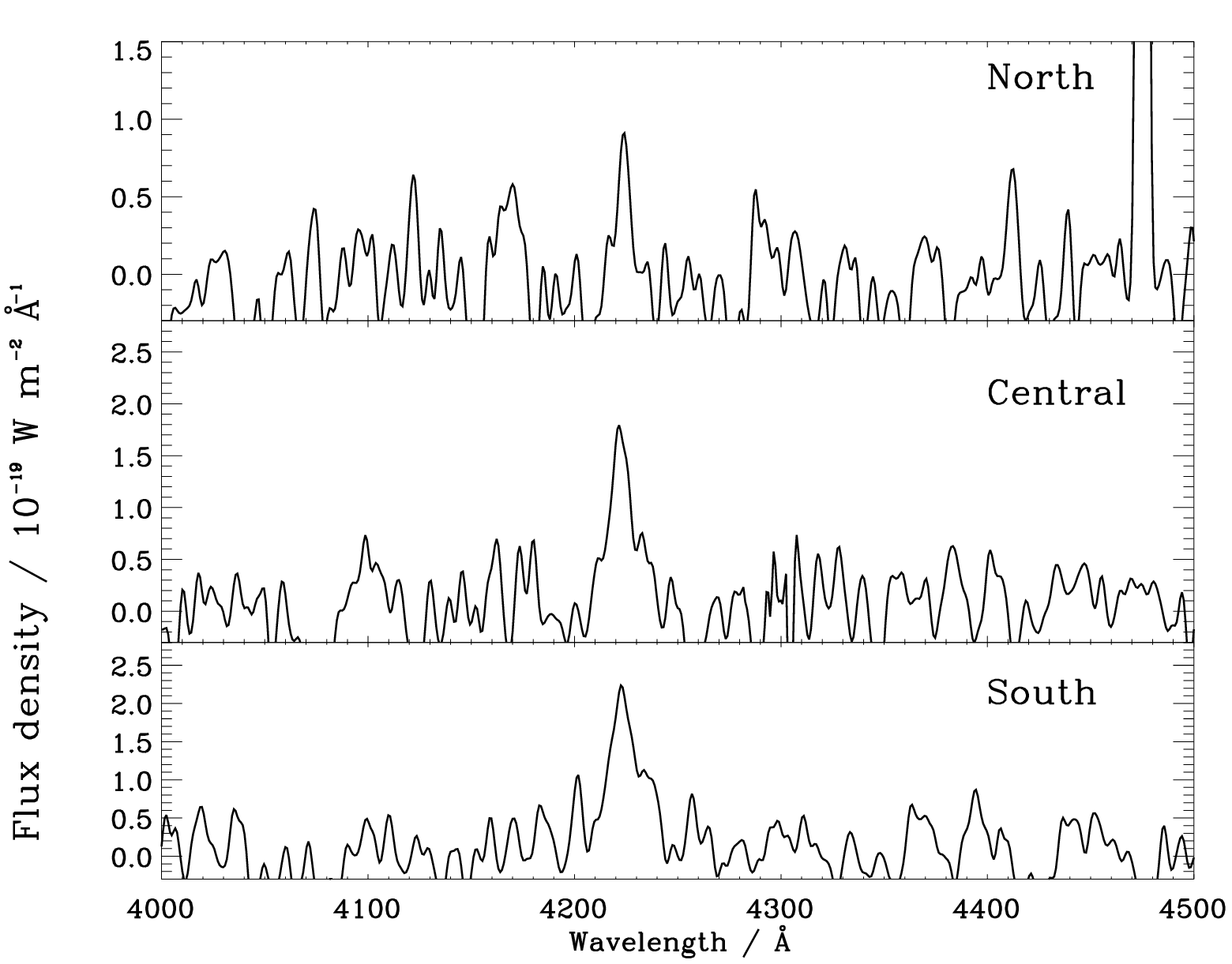}%
\caption{The left panel is the two dimensional optical spectrum centred one the
 position of the host galaxy and smoothed
over 3 pixels showing two clear and a third tentative spatially-distinct
Ly$\alpha$ blobs at the same wavelength. The right panel shows
spectra of the three blobs of emission: the central spectrum is from the
central region co-spatial with the $K$-band identification; the top and
bottom spectrum are from the northern and southern blob shown in
Fig.~\ref{F_overlays}. An
emission line is detected at 4222\AA\, and identified with
Ly$\alpha$ at $z = 2.47$ in all three blobs.}\label{F_o-spect}
\end{center}
\end{figure*}

Figure \ref{F_o-spect} shows the two-dimensional spectrum,
where three separate emissions lines can be identified,
associated with the three blobs present in the $g$-band image;
together with one-dimensional spectra extracted from  each
blob. We find an emission line at 4222 \AA\ in the central and
southern blobs, which we identify 
as Ly$\alpha$, corresponding to a redshift of $z = 2.47$. No
other lines (e.g. C {\sc iv}, He {\sc ii}, C {\sc iii}, etc.)
could be found associated with either blob. 
The combined flux from all of the Ly$\alpha$ blobs, after correcting
for the aperture and the edge of the filter,  can account for $\sim$
90\% of the flux seen in the $g$-band image. Furthermore, the spectrum
taken with GMOS-N, where only the central blob is positioned on the
slit, shows some faint continuum of the order \mbox{$g \sim$ 22}; which
corresponds to $\sim$ 30\% of the contribution of the central blob. 

This result is consistent with the $K$-band emission line being
H$\alpha$ at $z =2.49$; and not Mg{\sc ii} as it was first
thought. With this established, the tentative line in the $HK$
spectrum is now confirmed as the [O {\sc iii}]$\lambda$4959/5007
doublet at $\sim 1.74\,\mu{\rm m}$ (Fig. \ref{F_ir-spec}). We,
therefore, use the redshift determined form the [O {\sc
    iii}]$\lambda$5007 line  (i.e. z = 2.48), as the best estimate for
the redshift of our source. We chose not to trust the redshift from
the Ly$\alpha$ line as this could be affected by absorption and thus
its position is more uncertain, while the H$\alpha$ line is likely to
be contaminated with other emission lines, e.g. [N{\sc ii}]. The
emission line properties are summarised in Table \ref{T_spect}.

\begin{table}% TABLE 3 spectroscopy
\caption{Results (observed values) from the optical and
near-infrared spectroscopy of J0049+3510. The signal-to-noise ratio on the
$[$O {\sc iii}$]$ doublet was too low to allow
measurement of the line width.}\label{T_spect}
\begin{center}
\begin{tabular}{l r r r c}
\hline
\multicolumn{1}{c}{Line}&\multicolumn{1}{c}{$\lambda_{\rm Obs}$/} & \multicolumn{1}{c}{FWHM/}&\multicolumn{1}{c}{Line Flux}\\
& \multicolumn{1}{c}{\rm \AA}&\multicolumn{1}{c}{\rm km s$^{-1}$} & \multicolumn{1}{c}{$10^{-18}$\rm W~m$^{-2}$}\\
\hline
H$\alpha_{6563}$ & 22920 &$ 3300 \pm 500 $& $1.9 \pm 30$\%\\
$[$O {\sc iii}$]_{4959/5007}$ & 17340,17423 & \multicolumn{1}{c}{--}& $0.3 \pm 50$\%  \\
Ly$\alpha$ (north)&4224&$<650 \:\; $&$0.64 \pm 40$\%\\
Ly$\alpha$ (central)&4222&$1290\pm 220$&$2.4 \pm 10$\%\\
Ly$\alpha$ (south)&4222&$1550 \pm 280$&$3.4 \pm 15$\%\\
\hline
\end{tabular}
\end{center}
\end{table}

%\emph{Short paragraph here explaining what the de-convolved spatial
%sizes at northern and southern blobs are, and discussing
%de-convolved velocity widths (say in this what 1arcsec is in kpc}

The southern blob extends to $\sim 4$~arcsec ($\sim 33$~kpc)
from the central $K$-band identification, with the overall
extent of the Ly$\alpha$ emission, including the northern component,
spanning $\sim 10$~arcsec 
($\sim 83$~kpc), suggesting this object is embedded in a very
extended Ly$\alpha$ halo. The de-convolved velocity widths
of the Ly$\alpha$ lines extracted over the central and southern blob are very
similar, with  FWHM of $1290\pm 220$ and
$1550\pm280$~km~s$^{-1}$ (see Table~\ref{T_spect}). However, it is
clear from Fig.~\ref{F_o-spect} that there is evidence for
a broader wing redward of the main Ly$\alpha$ emission compared to the
blue side,  at least in the southern 
component and at lower signal-to-noise ratio, possibly due to dust
obscuration associated with the host galaxy extinguishing the LyA line,
in the central component. Broad red-wings are
indicative of inflows/outflows and have been seen in superwind
galaxies
\citep[e.g.][]{2002ApJ...576L..25A,2002ApJ...570...92D,2007arXiv0705.1494S}.
Alternatively, such a profile could also indicate the presence of a
neutral hydrogen absorbing screen similar to those seen around
radio galaxies of small projected linear size
\citep[e.g.][]{1997A&A...317..358V,2003MNRAS.338..263J,2004MNRAS.351.1109W,2006A&A...459...31B}.
Our near-infrared spectroscopy does not have the requisite
signal-to-noise ratio with which to determine the shape of the
H$\alpha$ line accurately. However, the fact that we see very
little sign of a broadened red component in H$\alpha$ leads us
to suggest that the broadened red wing component, when compared to the
blue side, is due to associated 
absorption in the blue wing of the Ly$\alpha$ emission.

To summarise, J004929.4+351025.7 is a radio-loud GPS source
at $z = 2.48$, reasonably bright in $J$, $H$ and $K$ but
invisible in both $I$ and $R$ down to $3\sigma$ limiting
magnitudes of $R = 23.9\,{\rm mag}$ and $I = 24.5\,{\rm mag}$.
It has a broad H$\alpha$ emission line with a very
weak [O {\sc iii}], and it lies within $83\,{\rm kpc}$
Ly$\alpha$ emitting halo.
%$\dagger$ surrounded/at the edge  by
%a large ($\sim 50$ {\rm  Mpc} $\dagger$ 60{\rm kpc})

\section{Discussion}\label{sec:discuss}

\subsection{Origin of the Ly$\alpha$ halo}
The radio SED of J0049+3510 (Fig. \ref{F_SED}) shows that
it is a powerful (L$_{1.4} = 7.5 \times
10^{25}$~W~Hz$^{-1}$~sr$^{-1}$) GPS radio source with a rest-frame
peak at $\nu \sim 4.2$~GHz.  Using equation~1 of
\cite{1998PASP..110..493O} to estimate the electron lifetime of
a typical GPS source we find that J0049+3510 is probably a
very young radio source, viewed $< 10^{4}\,{\rm yr}$ after
the jet-triggering event. We can also use the strong
anti-correlation between the turnover frequency and the
projected linear size of the source \citep[e.g.][]{1990A&A...231..333F}
to estimate the linear size of the
radio emission in J0049+3510. Using equation~4 in \cite{1998PASP..110..493O},
for a rest-frame turnover frequency of 4.2~GHz, we
find that the projected linear size would be $\sim 100 ~ \rm pc$,
which is consistent with jet hotspots advancing at $\sim 0.25 ~ \rm c$
\citep[as it has been measured in compact symetric radio
  sources;][]{1998A&A...337...69O} for $\sim 10^{3} ~ \rm yr$. The
inferred limit on the size is also consistent (at $z = 2.48$) with the 
$< 0.3$~arcsec limiting spatial
resolution of the 8.4~GHz radio observations.

If we were to consider a model in which
accretion activity, and hence the optical quasar nucleus,
was triggered at the same time as the
radio jet; then the ionising photons could have reached no
further than $\sim (1/0.25) \times 0.3 \simeq 1$~arcsec from the nucleus
and, as such, the quasar nucleus would have not be
able to ionise the Ly$\alpha$ halo that we observe.
Thus, if we expect  the Ly$\alpha$ halo to be ionised
by the central AGN, a longer living source of ionising photons
(e.g. a pre-existing accreting super-massive black hole and/or
a starburst) would be required to explain the presence of Ly$\alpha$
emission $\sim$ 40~kpc from the nucleus. Indeed, it would take a
minimum of $\sim 10^{5}$ years for the photons to reach the
furthermost parts of the Ly$\alpha$ halo. This is shorter than
the expected duty cycle of a typical AGN lifetime
\citep[e.g.][]{2003ApJ...597L.109M}, thus the Ly$\alpha$
halo could easily be photoionised by an obscured quasar, in place
before the current jet activity was triggered. 

\subsection{Estimating the extinction}
In order to investigate the reddening of the optical and
near-infrared SED, we have used a simple $\chi^2$-fitting technique to
estimate the amount of extinction. We used the SDSS composite quasar
template \citep{2001AJ....122..549V} combined with the AGN spectral
energy distribution from \citet{1994ApJS...95....1E}. To this, we
applied extinction corrections for three different types of dust
(Milky Way MW, Large Magellanic Cloud LMC, and Small Magellanic Cloud
SMC) as in \citet{1992ApJ...395..130P}.  We have neglected any contribution
from a host galaxy principally because any such component would peak
in the observed-frame SED around $J$ band, where the emission is
spatially unresolved, and secondly the host would have to be
unreasonably massive ($\gtsim 10^{12} {\rm M_{\odot}}$, 
corresponding to a $\gtsim 6 ~ \rm L_{*}$
elliptical after accounting for passive evolution) to contribute even
at the $\sim 10 \%$ level. Furthermore,  the $K$--$z$ relation for a
typical radio galaxy at $z = 2.48$ predicts a $K$ magnitude of $K =
19.1 \pm 0.5$ \citep{2001qhte.conf..333J, 2003MNRAS.339..173W}, $\sim$
1.5 mag fainter than our source. 

\begin{figure*}
\begin{center}
\includegraphics[height=0.96\textwidth, width=0.74\textwidth, angle=90]
{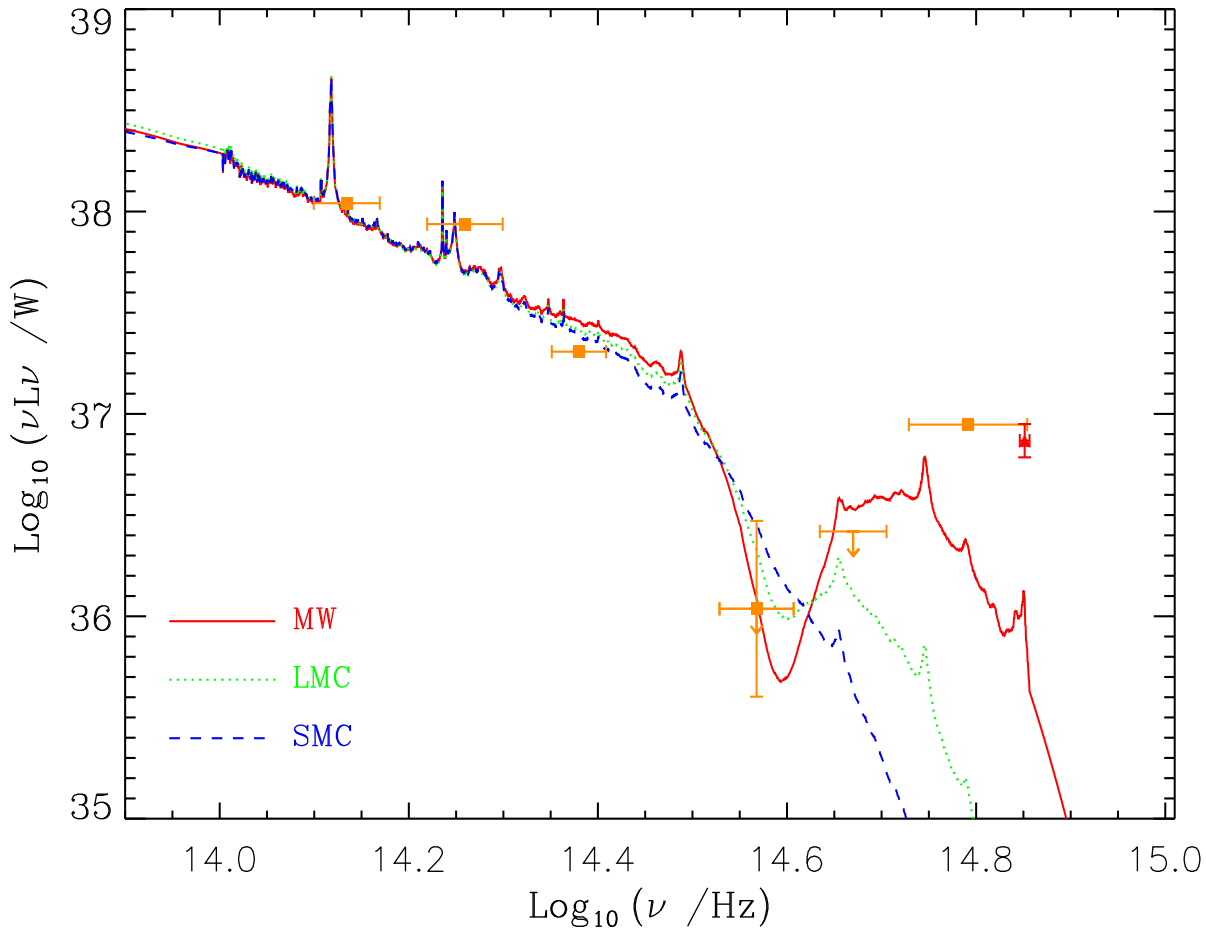}
\caption{Near-infrared and optical SED of J0049+3510.
The $KHJIR$ photometric data points, and $g$-band extended emission
detection (filled orange squares) 
suggest a reddened quasar ($I$ and $R$-band data points are $1.5
\sigma$ limits). We also show the  Ly$\alpha$ line observed on the
edge of the $g$-band in spectroscopy (filled red triangle). We also
plot the best-fit 
composite quasar spectra for different dust types (as different 
line styles and colours as labelled) yielding best-fit extinctions of
$A_V \approx 3$ in all cases. } \label{F_IRSED}
\end{center}
\end{figure*}

For the fitting we used our three near infrared photometry points plus the
$I-$band non-detection, which was included, following
\citet{2006MNRAS.370.1479M}, by assigning a flux density and an error both
equal to half the flux density limit (so $S_I = \sigma_I = 1.5
\sigma_{\rm obs}$). The best fits for the quasar extinction are
shown in Figure \ref{F_IRSED}. They are almost independent of
dust type and yield values for the rest-frame extinction of
$A_V \sim 3$. The  marginalised probability for the  extinction
  parameter for the 3 different types of dust is: \mbox{Av $=
    2.93^{+0.18}_{-0.24}$} for MW and LMC and \mbox{Av $=
    2.76^{+0.16}_{-0.24}$} for SMC. 

The Ly$\alpha$/H$\alpha$ line ratio might give extra information on
extinction on this object. However, as discussed in
\citet{1993ApJ...411...67E} it is not easy to obtain a reliable
intrinsic extinction value because of the effects of resonant
scattering of Ly$\alpha$. Therefore, we will only use this line ratio
as an independent verification of our fitting results. The line ratio
measured from the central blob is 1.8; similar to those found in
high--redshift radio galaxies (HzRG) by
\citet{1993ApJ...411...67E}. Moreover, if we assume an intrinsic
Ly$\alpha$/H$\alpha$=8 line ratio and apply an extinction of Av=3 with
a MW type dust, the Ly$\alpha$/H$\alpha$ line ratio we should observe
is 1.9, indicating that our fitting suggests a similar extinction to
the optical continuum of the quasar and the narrow emission lines. We
note that this level of reddening could be due to the edge of an
obscuring torus, or could be due to dust in the host galaxy, although
the latter is preferred as the narrow lines too appear to be
reddened. 

It is also plausible that the large diffuse line emission could be
caused by gas cooling on to the dark matter halo of the galaxy
\citep[e.g.][]{2001ApJ...562..605F, 2006ApJ...649...14D,
2006A&A...452L..23N, 2007astro.ph..3522S, Smithetal2008}. However, the
distribution of the ionised gas suggests a loosely biconical ionisation
front, as would be expected from a central AGN, with an
obscuring torus where the ionising photons can only escape
through the opening angle of the dusty torus. This biconical
structure is seen in other radio galaxies, and most
convincingly in Cygnus~A \citep{2003MNRAS.342..861T}. This
spatial distribution could also, in principle, be caused by a
large halo of absorbing gas around the galaxy as seen in many
radio galaxies with small angular sizes
\citep[e.g.][]{1997A&A...317..358V,2003MNRAS.338..263J,2004MNRAS.351.1109W,2001MNRAS.326.1563J}.
However, evidence suggests that such haloes may be roughly
spherical in nature \citep{2003MNRAS.338..263J}, and thus do not have a
preferred orientation, although such gas may account for the
observed broader redward component in the Ly$\alpha$
emission-line profile. Furthermore, there is no evidence of any
other extended emission around the central source apart from in
the biconical direction shown in Fig.~\ref{F_overlays}.

\section{Conclusions}\label{sec:conc}
During our search for $z>6$ radio-loud quasars we have
discovered an obscured quasar at $z =2.48$ embedded in a large
Ly$\alpha$ halo. The ionising source of the large-scale
Ly$\alpha$ halo is most likely to be the central AGN, with a
dusty torus or a dusty galaxy obscuring the AGN along our line-of-sight. The
dusty torus would also explain the structure of the Ly$\alpha$
emission-line halo, with the ionising photons essentially being
forced into a biconical distribution according to the opening
angle of the torus.

Using the peak of the radio spectral energy distribution to
estimate the age of the radio jet, we find that the radio
source has been active for $\sim 10^{3}$~years. assuming a
jet speed of $\sim 0.25c$. This implies that the accretion activity
must have started prior to the radio-jet triggering event,
suggesting that the triggering mechanism for the optical quasar
is different, or at least not coincident in time with, the triggering
of the radio jet.

Further work on J0049+3510, including integral field
spectroscopy, will allow us to map the emission line
profiles across the halo and examine the nature of the ionised
gas. Higher-resolution radio observations would also allow us to
test whether the ionising photons are being emitted in the same
direction as the young radio-jet.

Completion of the follow-up for the rest of the {\em `red'}
candidates quasars will quantify how common these type of
sources are, and hopefully find some genuine $z \gtsim 6.5$
radio-loud quasars.

\section*{acknowledgments}
FEBM was funded by {\sc sisco} research training network part
of European Commission's 5th Framework Improving Human
Potential programme. MJJ is supported by a Research Council UK
fellowship. The work of SC was performed under the auspices of
the U.S Department of Energy by the University of California,
Lawrence Livermore National Laboratory under contract
No.W-7405-ENG-48. SC acknowledges support for radio galaxy
studies at UC Merced, including the work reported here, with
the Hubble Space Telescope and Spitzer Space Telescope via NASA
grants HST \#10127, SST \#1264353, SST \#1265551, and SST
\#1279182. AM acknowledges support from grant \emph{AYA
2004$-$3136} from the Spanish Ministerio de Educaci\'on y
Ciencia. The United Kingdom Infrared Telescope is operated by
the Joint Astronomy Centre on behalf of the Science and
Technology Facilities Council of the U.K, some of the data
reported here were obtained as part of the UKIRT Service
Programme. The William Herschel Telescope and the Isaac Newton
Telescopes are operated on the island of La Palma by the Isaac
Newton Group in the Spanish Observatorio del Roque de los
Muchachos of the Instituto de Astrofísica de Canarias. The
Digitized Sky Surveys were produced at the Space Telescope
Science Institute under U.S. Government grant NAG W-2166. The
images of these surveys are based on photographic data obtained
using the Oschin Schmidt Telescope on Palomar Mountain and the
UK Schmidt Telescope. The plates were processed into the
present compressed digital form with the permission of these
institutions. Based on observations obtained at the Gemini
Observatory, which is operated by the 
Association of Universities for Research in Astronomy, Inc., under a
cooperative agreement 
with the NSF on behalf of the Gemini partnership: the National Science
Foundation (United 
States), the Science and Technology Facilities Council (United
Kingdom), the 
National Research Council (Canada), CONICYT (Chile), the Australian
Research Council 
(Australia), Ministério da Ciência e Tecnologia (Brazil) and SECYT
(Argentina).

%\bibliographystyle{D:/Libraries/LaTeX_libraries/class_files/apj}
%\bibliography{D:/Libraries/LaTeX_libraries/class_files/aamnem99,D:/Libraries/LaTeX_libraries/references_db/reference_database}

\end{document}